\documentstyle[11pt,paspconf]{article}

\input{psfig}  
\begin{document}

\title{New Dwarf Galaxies in the IC342/Maffei Group}  

\author{W. K. Huchtmeier} 
\affil{Max-Planck-Institut f\"{u}r Radioastronomie, 
       Bonn, D-53121, Germany} 

\author{I. D. Karachentsev}
\affil{Special Astrophysical Observatory (SAO), N. Arkhys, Russia } 

\author{V. E. Karachentseva}
\affil{Astronomical Observatory of Kiev State University, Kiev, Ukraine}

\begin{abstract}
This is a report on the detection  of HI-emission from
three 'new' dwarf galaxies Perseus\,A, Perseus\,B, and Camelopardalis\,D
in the IC342/Maffei group of galaxies and of Draco\,A. 
The actual number of (probable) member galaxies of this group increases
to 19 galaxies.  Its velocity dispersion is 86 km\,s$^{-1}$.
With a distance of 2.2$\pm$0.5 Mpc this group is the nearest to the
Local Group and might have considerable dynamical influence on the Local
Group.
\end{abstract}


\section{Introduction}

Due to its position within the zone of avoidance between the Andromeda
region of the Local Group and the M81 group the IC342/Maffei group has
been recognized only lately as a group.
Since 1994 a great number of dwarf galaxies have been discovered in this area
with growing interest for galaxies in the zone of avoidance.
There have been blind HI surveys and optical searches for galaxies in
the area.
Recent discoveries of Dwingeloo 1 (Kraan-Korteweg et al. 1994,
Huchtmeier et al. 1995), Dwingeloo 2 (Burton et al. 1996), Cas 1 
(Huchtmeier et al. 1995), MB1 and 2 (McCall and Buta 1995, McCall et al.
1995), Cam B (Huchtmeier et al. 1997), MB3 (McCall and Buta 1997)
have increased the number of known galaxies in this group considerably.
Here we report HI-detection of the dwarf galaxies Camelopardalis D,
Perseus\,A and B, and of Draco A 
 in the area of the M81 group. \footnote{Possibly more detections 
have been reported by Rivers 1998.}  
The IC342/Maffei group is the nearest group outside the Local Group
with a photometric distance of 2.2 Mpc.

A new list of candidates of nearby dwarf galaxies from the sky survey
of  surface brightness dwarf galaxies based on the POSS II and
ESO/SERC films has been searched for HI emission with the 100-m
radiotelescope at Effelsberg. So far two lists of the Karachentsev
survey have been published (Karachentseva and Karachentsev 1998, 
Karachentseva et al. 1999), HI observations of some of these new dwarf
galaxies have been reported (Huchtmeier et al. 1997, Huchtmeier et al.
1999). 
Among those newly discovered galaxies three are situated in the
IC342/Maffei group according to their position and radial velocity. 

\section{Observations}

\begin{figure}   
\psfig{figure=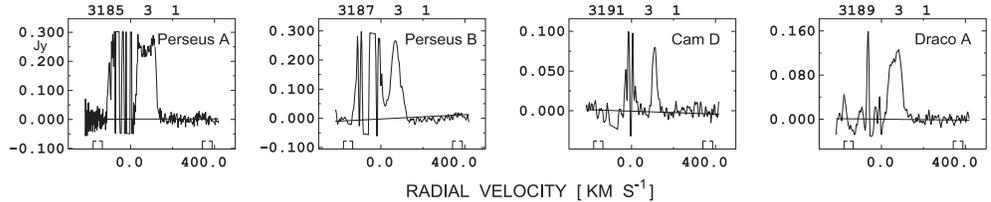,width=13cm} 
\caption{HI profiles of the 4 'new' dwarf galaxies in the
area of the IC342/Maffei group observed with the 100-m radio telescope
at Effelsberg, from left to right : Perseus A, Perseus B, Camelopardalis
D, and Draco 1. All profiles but Perseus A  have been Hanning smoothed.
The channel spacing is 2.6 km\,s$^{-1}$.  
} 
\end{figure} 

%Table 1  
\footnotesize 
\begin{table}  
\caption{HI parameters of the 'new' dwarf galaxies in the area of the 
IC342/Maffei Group } 
%\newline 
\begin{tabular}{|llcrrrl|} \hline
\multicolumn{1}{|l}{Name}&  
\multicolumn{2}{l}{R.A. (1950.0) Dec.}&
%\multicolumn{1}{r}{$a  b$}&
%\multicolumn{1}{l}{Type}&
%\multicolumn{1}{c}{S.B.}&
%\multicolumn{1}{c}{optical}&
\multicolumn{1}{c}{velocity}& 
\multicolumn{1}{c}{HI-flux}& 
\multicolumn{1}{c}{S$_{max}$}&   
\multicolumn{1}{c|}{line-width}  
 \\ \hline   
\multicolumn{1}{|c}{ }&
\multicolumn{1}{c}{ }&
\multicolumn{1}{c}{}&
\multicolumn{1}{c}{km\,s$^{-1}$ }&
\multicolumn{1}{c}{Jy km\,s$^{-1}$ }&
\multicolumn{1}{c}{Jy }&
\multicolumn{1}{c|}{50\% 20\% }  
 \\ \hline   
&&&&&&  \\
Perseus\,A & 02 21 03.7 &  55 37 09& 75$\pm$1 & 24.2 & 0.29$\pm$0.025& 94 104
\\
Perseus\,B & 02 23 51.3 & 57 15 50& 72$\pm$2 & 14.0 & 0.26$\pm$0.016& 46 86
\\
Cam\,D & 05 53 23.1 &  73 25 24& 111$\pm$2& 2.1& 0.08$\pm$0.008& 23 34 \\
Draco\,A & 12 11 42.2  & 66 22 12& 66$\pm$1& 8.1& 0.13$\pm$0.012& 68 81 \\
&&&&&&  \\   
\hline 
\end{tabular} 
\end{table} 

\rm 
\normalsize 

Observations were performed with the 100-m radio telescope at Effelsberg
which has a half power beam width (HPBW) of 9.3' at the wavelength of 21-cm.
Observations have been obtained in the total power mode  combining the
on-source position with a reference field. A bandwidths of 3.125 MHz was split
into four channels yielding a channels spacing of 12.2 kHz and a
resolution of 3.1 km\,s$^{-1}$ (or 5.1 km\,s$^{-1}$ after Hanning
smoothing). 
For all galaxies four additional 
 positions a beam width off the central position in R.A. and Dec. 
 have been observed to check
for extend of the HI. The HI emission was centered to the optical
positions and nearly not extended compared to the HPBW.
The profiles are shown in Fig. 1, the observed HI parameters in Table 1. 
Apart from Cam\,D all profiles seem partially confused by local HI 
which is seen best for Perseus\,A and B. 

\footnotesize 
\begin{table}[t]  
\caption{Galaxies in the IC342/Maffei Group } 
%\newline 
\begin{tabular}{|llcrlrc|} \hline
\multicolumn{1}{|l}{Name}&  
\multicolumn{1}{l}{Type}&
\multicolumn{1}{c}{optical}&
\multicolumn{1}{c}{velocity}& 
\multicolumn{1}{c}{$M_B$}& 
\multicolumn{1}{c}{$M_{HI}$}&   
\multicolumn{1}{c|}{$M_{HI}/L_{B}$}  
 \\ \hline   
\multicolumn{1}{|c}{ }&
\multicolumn{1}{c}{ }&
\multicolumn{1}{c}{diameter}&
\multicolumn{1}{c}{v$_{0}$ }&
\multicolumn{1}{c}{ }&
\multicolumn{1}{c}{ }&
\multicolumn{1}{c|}{ }  
 \\ \hline   
\multicolumn{1}{|c}{ }&
\multicolumn{1}{c}{ }&
\multicolumn{1}{c}{arc min}&
\multicolumn{1}{c}{$km\,s^{-1}$ }&
\multicolumn{1}{c}{ }&
\multicolumn{1}{c}{10$^{7}$M$_{\odot}$ }&
\multicolumn{1}{c|}{ }  
 \\ \hline
\multicolumn{1}{|c}{1}&
\multicolumn{1}{c}{2}&
\multicolumn{1}{c}{3}&
\multicolumn{1}{c}{4}&
\multicolumn{1}{c}{5}&
\multicolumn{1}{c}{6}&
\multicolumn{1}{c|}{7} 
 \\ \hline 
&&&&&&    \\
IC342  & Scd  & 21.4 20.9 & 229 & -20.3 & 355 & 0.34  \\  
Maffei I&  E  &  3.3 1.7  & 224 & -19.5 &     &       \\ 
Maffei II& Sbc& 5.8  1.6  & 209 & -19.5 & 35  & 0.04  \\ 
Dw1    & SBcd  &  4.2 0.3  &309 & -17.5 &  24 & 0.95  \\ 
NGC 1569& IBm &  3.6 1.8  &  87 & -17.1 &  8 & 0.05  \\ 
&&&&&&  \\ 
NGC 1560& Sd  & 11.6   0.8 & 164& -15.5 & 27  & 1.28  \\ 
UGCA105 & Im? &  5.5 3.5  & 264 & -15.3 & 20  & 1.12  \\ 
UGCA 92 & Im? & 2.0 1.0 &   66 &  -14.0 &  6 & 0.49  \\ 
UGCA 86 & Im? & 0.8 0.7 &  262 &  -12.7 &  8 & 0.93  \\ 
Cas 1  &  dIm & 1.9 1.6 &  264 &  -14.8 & 5 & 0.39  \\ 
&&&&&&  \\ 
MB1    &  Sd  & 6.0 1.0 &  398 &  -12.5 & 2.2 & 1.44  \\ 
Dw2    & Im &   6.4     &  291 &  -13.1 & 4.1 & 1.49  \\ 
Cam B  & Im &  2.2  1.1 &  264 &  -12.1 & 0.3 & 0.28 \\ 
Cam D  & Im &  0.6 0.5 &  278 &   -11.0 & 0.12 & 0.29 \\ 
Perseus A  & Im &  1.4 0.8 &  288 &   -12.9 & 1.4 & 0.58 \\ 
&&&&&& \\  
Perseus B  & Im &  1.7 0.5 &  283 &   -12.7  & 1.2 & 0.66 \\ 
Cam C  & dIm & 1.8 0.4 &  151 &          &      &      \\ 
Cam A  & dSph & 3.7 2.1 &      &          &      &      \\ 
MB3    & dSph& 1.6 0.5 &      &          &      &      \\ 
&&&&&& \\ 
\hline  
\end{tabular}
\end{table}

\rm 
\normalsize

\begin{figure}[t]    
\psfig{figure=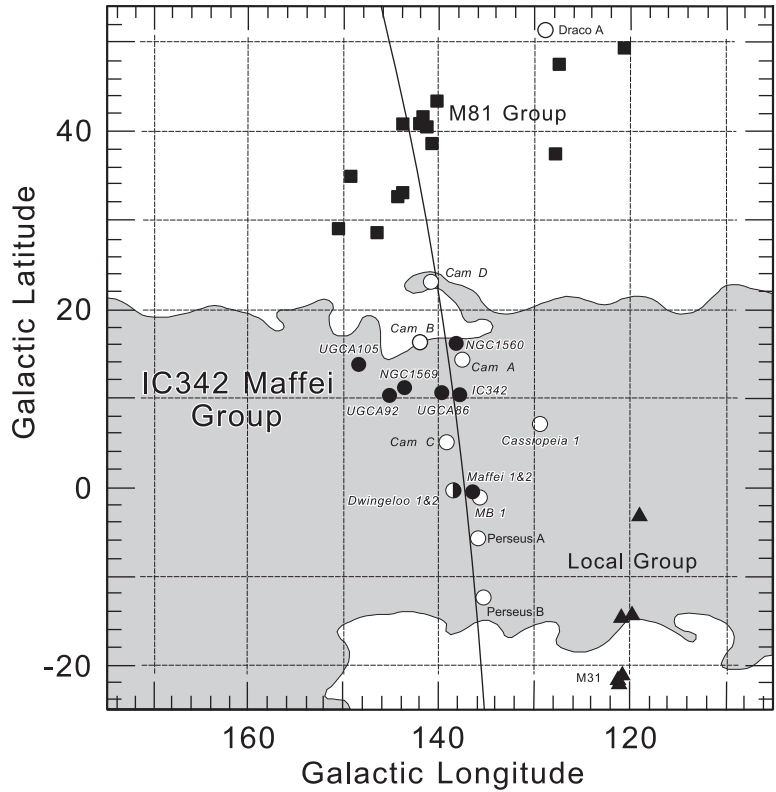,width=11cm} 
\caption{The distribution of galaxies in the
area of the IC342/Maffei group (indicated by circles) along the
supergalactic equator in galactic coordinates. Galaxies detected
since 1994 are given by open circles.} 
\end{figure}

\section{Discussion} 

The clustering of galaxies around IC342 and Maffei 1 (Fig. 2) within the
zone of avoidance along the supergalactic equator in addition to the 
similar corrected radial velocities
(Table 1) suggests strongly a typical group of galaxies. In the past 
greatly different distances have been quoted for this group , see
discussion by Krismer et al. (1995) and by McCall and Buta (1997). 
In recent years photometric distances have been derived for 10 galaxies
in this group (Karachentsev and Tikhonov 1993, 1994, Karachentsev et al.
1997, and unpublished
work). These distances agree quite well with each other and yield a 
distance of 2.2$\pm0.5$ Mpc for the IC342/Maffei group.
At such a close distance it might have played a significant role in the
dynamical evolution of the Local Group (McCall 1986, 1989, Zheng et al.
1991,
Valtonen et al. 1993, Peebles 1994).

\acknowledgments

The NED database is supported at IPAC by NASA. 

This work has been partially supported by the Deutsche
Forschungsgemeinschaft (DFG) under project no. 436 RUS 113/470/0.


\small  
\begin{references}
\reference Burton, W. B., Verheijen, M. A. W., Kraan-Korteweg, R. C.,
	   Henning, P. A. 1996, A\&A, 309, 687 
\reference Huchtmeier, W. K., Lercher, G., Seeberger, R., Saurer, W.,
	   Weinberger, R. 1995, A\&A, 293, L33 
\reference Huchtmeier, W. K., Karachentsev, I. D., Karachentseva, V. E.
           1997, A\&A, 322, 375 
\reference Karachentseva, V.E., Karachentsev, I.D. 1998, A\&A Suppl., 127, 409
\reference Karachentseva, V. E., Karachentsev, I. D., Richter, G.M.
           1999, A\&A Suppl., 135, 221 
\reference Karachentsev, I. D., Drozdovsky, I., Kajsin, S., Takalo, L.
    O., Hein\"{a}m\"{a}ki, P., Valtonen, M. 1997, A\&A Suppl. 124, 559  
\reference Karachentsev, I. D., Makarov, D. A. 1996, \aj, 111, 794 
\reference Karachentsev, I. D., Tikhonov, N. A. 1994, A\&A, 286, 718 
\reference Karachentsev, I. D., Tikhonov, N. A. 1993, A\&A Suppl. 100,
227  
\reference Kraan-Korteweg, R. C., Loan, A. J., Burton, W. B., Lahav, O.,
	   Ferguson, H. C., Henning, P. A., Lynden-Bell, D. 1994,
	   Nature, 372, 77 
\reference Krismer, M., Tully, R. B., Gioia, I. M. 1995, \aj, 110, 1584
\reference Lynden-Bell, D., \& Wood, R.  1968, \mnras, 138, 495
\reference McCall, M. L. 1986, JRASC, 80, 271 
\reference McCall, M. L. 1989, \aj, 97, 1341 
\reference McCall, M. L., Buta, R. J. 1995, \aj, 109, 2460 
\reference McCall, M. L., Buta, R. J., Huchtmeier, W. K. 1995, IAU Circ.
           No. 6159  
\reference Peebles, P. J. E. 1994, \apj, 429, 43 
\reference Rivers, A. J., 1998, AAS, 193, 28.05  
\reference Valtonen, M. J., Byrd, G.G., McCall, M. L., Innanen, K. A.
           1993, \aj, 105, 886 
\reference Zheng, J. Q., Valtonen, M. J., Byrd, G. G., 1991 A\&A 247, 20

\end{references}
\end{document}